\documentclass[aps,fleqn,preprint]{revtex4-1}

\usepackage{amsmath}
\usepackage{graphicx}
\usepackage{epsfig}
\usepackage{subfigure}
\usepackage{amsfonts}
\usepackage{amssymb}
\usepackage{amsthm}
\usepackage{newlfont}
\usepackage{epstopdf}

\usepackage[utf8]{inputenc}

\usepackage{dcolumn}
\usepackage{bm}

\usepackage[section] {placeins}

\begin{document}

\title{Equilibrium thermodynamic properties of binary hard-sphere mixtures from integral equation theory }

\author{Banzragch Tsednee}
\author{Tsogbayar Tsednee}
\author{Tsookhuu Khinayat}
\affiliation{Institute of Physics and Technology, Mongolian Academy of Sciences, Peace Ave 54B, 13330 Ulaanbaatar, Mongolia }

\begin{abstract}

The binary additive hard-sphere mixtures have been studied by the Ornstein-Zernike integral equation coupled with the Martynov-Sarkisov (MS) closure approximation. Virial equation of state is computed in the MS approximation. The excess chemical potential for the mixture is evaluated with a closed-form expression based on correlation functions. The excess Helmholtz free energy is obtained using the Euler relation of thermodynamics. Moreover, these thermodynamic quantities are obtained by the Boubl\'{i}k-Mansoori-Carnahan-Starling-Leland (BMCSL) formulas. Our findings for pressure and excess chemical potential for a number of binary sets of the mixtures from the MS approximation show good agreements with those obtained by the BMCSL formulas and available data in literature, having a maximum deviation of $5\%$ for a packing fraction up to 0.5. The maximum deviation of the excess free energy obtained for the mixtures is shown to be $\sim 16\%$ for a packing fraction of 0.5.  To our knowledge, this work presents an initial calculation of an excess chemical potential of the system in the MS approximation.

\end{abstract}

\pacs{Valid PACS appear here}
                             

\maketitle

\section{Introduction}

The hard-sphere (HS) model plays a major role in the development of the modern theory of liquids \cite{Hansen_06}. 
This system is often used as a standard reference system in the perturbation approach of the liquids for studying the liquid-state properties \cite{McQuarre_76, Hansen_06}.    
Mixtures of the HSs of different size distributions can be employed to model the colloidal suspensions, which play an important role in chemical and bioengineering fields~\cite{Heno91, Pago01}. 

The binary HS mixture is the simplest model of the multicomponent systems. Computer simulations for its thermodynamic and structural properties began in the mid-1960s. Alder~\cite{Alder64} used the molecular dynamic (MD) simulation to solve the equation of state (EOS) for the binary mixture. Rotenberg~\cite{Rotenberg65} performed the Monte-Carlo (MS) EOS calculation for a mixture of HSs. An alternative approach to study a liquid is the integral equation (IE) method, in which the IE combined with an approximate closure is solved to predict the structure and to obtain thermodynamic properties. Lebowitz \cite{Lebowitz64} obtained an exact solution of the PY integral equation for a multicomponent additive mixture. Mansoori {\it et al.}~\cite{Mansoori71} investigated the EOS for the mixture of the HSs using the MC and MD simulations, and proposed an analytical expression for the EOS using the solution the PY integral equation ~\cite{PY58}. Baro{\v s}ov{\'a} {\it et al.}~\cite{Barosova96} applied a test particle insertion method to obtain the excess chemical potential of the binary HS mixtures. Santos and his co-workers have investigated extensively thermodynamic properties and structural nature for multicomponent HS fluids in terms of the analytical approaches from the IE method~\cite{Santos13, Heyes18} and MD simulation \cite{Santos20}. Moreover, Ballone {\it et al.}~\cite{Ballone86} tested the PY approximation for the HS mixtures. Schmidt~\cite{Schmidt92} and Malijevsk{\'y} {\it et al.}~\cite{Malijevsky97} applied the Martynov-Sarkisov (MS)~\cite{Martynov83} closure for the binary mixtures as well. 

In this work our purpose is to obtain thermodynamic properties for binary additive HS mixture using the Ornstein-Zernike equation method coupled with the MS closure approximation.  
We will compute pressure using a virial route in the MS approximation. The excess chemical potential has been computed by an approximate expression based on correlation functions. The excess free energy will be obtained from thermodynamic relation. Moreover, we will compute these thermodynamic quantities using the Boubl\'{i}k-Mansoori-Carnahan-Starling-Leland (BMCSL)~\cite{Boublik70, Mansoori71} formulas, which are based on closed-form formulas obtained in the virial and compressibility routes from the PY integral equation theory.  
We will compare some of our findings for pressure, excess free energy and chemical potential with those obtained by MC method \cite{Barosova96} and MD simulation~\cite{Heyes18} and will present other new findings as well. To our knowledge, this is the first attempt to calculate the excess chemical potential for the mixture in the MS approximation. Therefore, we hope that it can be considered as a contribution in this field.  

The organization of this paper is as follows. In the first section the Introduction is presented. The following two sections discuss about theoretical formulations of the IE method for multicomponent mixture and thermodynamic quantities, which we will compute, and of our numerical results and their discussions, respectively. Then a conclusion follows.


\section{Theory}
\subsection{Integral equation theory}	

For binary system with the total number density $\rho$, the Ornstein-Zernike (OZ) equation, which establishes a relation between the total correlation function and the direct correlation function, has a form   
\begin{eqnarray}\label{ssoz}
	h_{ij}(\mathbf{r}) = c_{ij}(\mathbf{r}) + \rho \sum^{2}_{k=1} x_{k} \int d\mathbf{r}' c_{ik}(\mathbf{r}-\mathbf{r}') h_{kj}(\mathbf{r}') 
\end{eqnarray}
where $h_{ij}(r)$ and $c_{ij}(r)$ are the total and direct correlation functions, respectively, and $x_{i}= \rho_{i}/\rho$ is the mole fraction for the component $i$ with $\sum_{i=1}^{2}x_{i}=1$.

Since the OZ equation (\ref{ssoz}) contains the unknown correlation functions, it cannot be solved directly. In order to solve this equation an another equation, which is called a {\it closure relation} must be introduced, that couples the total and direct correlation functions with the pair interaction potential. 
A general closure equation for the mixture may be written in the form
\begin{equation}\label{closure}
	h_{ij}(r) = \exp[-\beta u_{ij}(r) + \gamma_{ij}(r) + B_{ij}(r)] - 1 \qquad (i,j = 1,2). 
\end{equation}  
Here $u_{ij}(r)$ is an interaction potential between particles in the system; $\gamma_{ij} \equiv h_{ij} - c_{ij}$ is an indirect correlation function; $B_{ij}(r)$ is the bridge function; $\beta = 1/k_{\mathrm{B}}T$, and $k_{\mathrm{B}}$ is the Boltzmann constant and $T$ is the temperature for the system. Then the OZ equation (\ref{ssoz}) and closure equation (\ref{closure}) is solved in self-consistent manner, and the total and direct correlation functions can be found numerically. 

For HS mixture, a form of an interaction potential in this work is given by
\begin{equation}\label{uij}
	u_{ij}(r) = 
	\begin{cases}
		\infty, & \quad r<\sigma_{ij}, \\
		0, & \quad r\geq \sigma_{ij},
	\end{cases}
\end{equation}
with $\sigma_{ij} = \frac{1}{2}(\sigma_{i} + \sigma_{j})$ and $\sigma_{ii} = \sigma_{i}$ is the hard sphere diameter for the component $i$.   

Since the exact analytic expression for the bridge function is not known in the liquid theories,  its approximated form has been mostly employed.  The form of the Martynov-Sarkisov~\cite{Martynov83} approximation, which we are using in this work, has a following form:
\begin{equation}\label{MS}
	B_{ij} (r) = (1+ 2\gamma_{ij})^{1/2} - \gamma_{ij} -1. 
\end{equation}

\subsection{Thermodynamic properties}

Once a solution of the integral equation is obtained, thermodynamic properties can be evaluated analytically as follows. 

\subsubsection{Pressure}

For the binary HS mixture, in the MS approximation we compute a pressure using a virial (v) route, in which pressure is evaluated via 
\begin{equation}\label{vir_pr}
	\frac{\beta p^{\mathrm{MS}}_{v}}{\rho} = 1 + \frac{2\pi}{3}\rho \sum^{2}_{i,j=1} x_{i}x_{j} \sigma^{3}_{ij} g_{ij}(\sigma_{ij}),  
\end{equation} 
where $g_{ij}(\sigma_{ij})$ are the contact values of the radial distribution functions defined as $g_{ij} (r) = h_{ij}(r) + 1$.  

We will also use the BMCSL pressure expression \cite{Santos20}, which is obtained as an interpolation between pressures from virial and compressibility routes in the Percus-Yevick approximation:  
\begin{eqnarray}\label{pr_bmcsl}
 	\frac{\beta p^{\mathrm{BMCSL}}}{\rho} =  \frac{1}{1-\eta} + \frac{3 \eta}{(1-\eta)^{2}} \frac{m_{2}}{m_{3}} + \frac{\eta^{2}(3-\eta)}{(1-\eta)^{3}} \frac{m^{3}_{2}}{m^{2}_{3}}.
\end{eqnarray}
In formula (\ref{pr_bmcsl}) $m_{n}$ is the reduced moments given by $m_{n}\equiv M_{n}/M_{1}^{n}$, where $M_{n}$ is the $n$th moment of the diameter distribution, which is given by  
\begin{eqnarray}\label{nth_mom}
	 M_{n} \equiv \langle \sigma^{2} \rangle = \sum^{2}_{i=1} x_{i} \sigma^{n}_{i},
\end{eqnarray}
where $\eta = \pi \rho M_{3}/6$ is a parking fraction, and $x_{i}$ is a mole fraction of the component $i$.

\subsubsection{An excess chemical potential}

To calculate the excess (e) chemical potential for the component $i$ in the mixtures in the MS approximation, we use a following closed-form expression based on the correlation functions \cite{Tsog19, Banz22, Banz22_num}
\begin{eqnarray}\label{excp}
\beta \mu^{e}_{i} \approx \sum_{j=1}^{2} \rho_{j} \int d\mathbf{r} \Big[ \Big(\frac{1}{2}h^{2}_{ij} - c_{ij} - \frac{1}{2}h_{ij} c_{ij} \Big) 
 + \Big(1 + \frac{2h_{ij}}{3}  \Big) B_{ij} \Big].
\end{eqnarray}
Note that this approximated expression (\ref{excp}) for the excess chemical potential can be used for any bridge functions since it does not require the explicit forms of them, and a derivation of this expression can be found in the Appendix.  

Similarly to the BMCSL expression for pressure, one can employ the BMCSL expression for an excess chemical potential, which has a following form \cite{Santos20} 
\begin{eqnarray}\label{excp_bmcsl}
 \beta \mu^{e^{\mathrm{BMCSL}}}_{i} & = & - \ln(1-\eta) + \frac{3\eta}{(1-\eta)} \frac{m_{2}}{m_{3}} \frac{\sigma_{i}}{M_{1}} + \Big[\frac{3\eta}{1-\eta} \frac{m_{2}}{m_{3}} + X_{2}(\eta) \frac{m^{3}_{2}}{m^{2}_{3}} \Big] \frac{\sigma^{2}_{i}}{M_{2}} \\ \nonumber
&& +  \Big[\frac{\eta}{1-\eta} + \frac{3 \eta^{2}}{(1-\eta)^{2}} \frac{m_{2}}{m_{3}}  + X_{3}(\eta) \frac{m^{3}_{2}}{m^{2}_{3}} \Big] \frac{\sigma^{3}_{i}}{M_{3}}
\end{eqnarray}
where $X_{2}(\eta) = 3\eta/(1-\eta)^{2} + 3\ln(1-\eta)$ and $X_{2}(\eta) = (5\eta^{2}-2\eta -\eta^{3})/(1-\eta)^{3} - 2\ln(1-\eta)$.

\subsubsection{An excess free energy}

Once we have the values of the compressibility factor $Z\equiv \beta p^{\mathrm{MS}}_{v}/\rho$ and the excess chemical potential $\beta \mu^{e}_{i}$,  we can compute the excess Helmholtz free energy per particle using a following Euler equation of thermodynamics
\begin{equation}\label{hfe}
	\beta a^{e} = \sum_{i=1}^{2} x_{i} \beta \mu^{e}_{i} - Z + 1. 	 
\end{equation}

We use the BMCSL formula for an excess free energy per particle, which is given in Ref.~ \cite{Santos20}: 
\begin{equation}\label{hfe_bmcsl}
	\beta a^{e^{\mathrm{BMCSL}}} = -\ln(1-\eta) + \frac{3\eta}{1-\eta} \frac{m_{2}}{m_{3}} + \Big[\frac{\eta}{(1-\eta)^{2}} + \ln(1-\eta) \Big] \frac{m^{3}_{2}}{m^{2}_{3}}. 
\end{equation}

\section{Results and Discussion}		

In our calculation we chose a component 1 with a diameter $\sigma_{1}$ as a reference particle, which is a larger component of binary HS mixture, that is, $(\sigma_{1} > \sigma_{2})$. A simple Picard iterative method for solving the OZ equation (\ref{ssoz}) was employed, and the numerical tolerance for the root-mean-squared residual of the indirect correlation functions during successive was set at $10^{-8}\sigma_{1}$. For all calculations a number of grid points is $2^{15}$ and a length parameter is $16\sigma_{1}$.

We first did numerical calculations for binary mixtures at values of $\eta = 0.15, 0.25, 0.35$ and $0.45$, and for $\sigma_{2}/\sigma_{1} = 0.5$ and $x_{1}=0.5$ values. Table 1 shows numerical result for the EOSs and their comparisons with an accurate MD values \cite{Heyes18}. Values in columns 2 and 3 are obtained with equations (\ref{vir_pr}) and (\ref{pr_bmcsl}), respectively. For each density, the absolute relative deviation percent (ARD\%) error for the MS and BMCSL values is obtained by means of accurate MD values in column 4, and is shown in parenthesis. From this comparison, the BMCSL formula presents quite accurate value for both low and high densities. However, in the MS approximation, as density increases, the ARD value increases. 
\begin{table}[h]
	\caption{Values of pressure from the MS and BMCSL aproximations at different values of a packing fraction $\eta$, and $x_{1} = 0.5$, and $\sigma_{2}/\sigma_{1} = 0.5$ and MD results~\cite{Heyes18}. The ARD value for each density is shown in parenthesis $(\%)$. }
	\begin{center}
		{\scriptsize
			\begin{tabular}{c@{\hspace{2mm}}c@{\hspace{2mm}}c@{\hspace{2mm}}c@{\hspace{2mm}}c@{\hspace{2mm}}c@{\hspace{2mm}}c@{\hspace{2mm}}c@{\hspace{2mm}}c@{\hspace{2mm}}c@{\hspace{2mm}}c@{\hspace{2mm}}c@{\hspace{2mm}}c@{\hspace{2mm}}c@{\hspace{2mm}} }
				\hline\hline
				$\eta$ & $\beta p^{\mathrm{MS}}_{v}/\rho$  &  $\beta p^{\mathrm{BMCSL}}/\rho$  & MD~\cite{Heyes18} \\
				\hline
				0.15 & 1.8 (1.3) & 1.7761 (0.068) & 1.7773   \\
				0.25 & 2.7 (2.3) & 2.7588 (0.195) & 2.7642 \\
				0.35 & 4.5 (0.8) & 4.5216 (0.322) & 4.5362 \\
				0.45 & 7.8 (2.0) & 7.9320 (0.381) & 7.9623 \\
				\hline\hline
		\end{tabular} }
	\end{center}
\end{table}  

In Table 2 we compared numerical values of the excess chemical potential from the MS and BMCSL approximations for each component with the MD values \cite{Heyes18}. The MS and  BMCSL values were obtained with the formulas (\ref{excp}) and (\ref{excp_bmcsl}), respectively. The corresponding ARD\% value for each value of $\eta$ is also shown in parenthesis. 
\begin{table}[h]
	\caption{ The same as shown in Table 1, but for the excess chemical potential.}
	\begin{center}
		{\scriptsize
			\begin{tabular}{c@{\hspace{2mm}}c@{\hspace{2mm}}c@{\hspace{2mm}}c@{\hspace{2mm}}c@{\hspace{2mm}}c@{\hspace{2mm}}c@{\hspace{2mm}}c@{\hspace{2mm}}c@{\hspace{2mm}}c@{\hspace{2mm}}c@{\hspace{2mm}}c@{\hspace{2mm}}c@{\hspace{2mm}}c@{\hspace{2mm}} }
				\hline\hline
				$\eta$ & $\beta \mu^{e}_{1}$ (MS) & $\beta \mu^{e}_{1}$ (BMCSL) & MD~\cite{Heyes18} & & & $\beta \mu^{e}_{2}$ (MS) & $\beta \mu^{e}_{2}$ (BMCSL) & MD~\cite{Heyes18} \\
				\hline
				0.15 & 2.1 (0.3) & 2.0953 (0.555) & 2.1070 & & & 0.7 (5.1) & 0.7338 (0.542) & 0.7378  \\
				0.25 & 4.5 (0.7) & 4.5279 (0.095) & 4.5322 & & & 1.5 (2.2) & 1.4736 (0.354) & 1.4684  \\
				0.35 & 8.6 (0.8) & 8.6287 (0.472) & 8.6696 & & &  2.6 (0.3) & 2.5819 (0.405) & 2.5924  \\
				0.45 & 16 (0.9) & 16.1394 (0.006)& 16.1404 & & & 4.5 (2.0) & 4.3843 (0.581) & 4.4099 \\
				\hline\hline
		\end{tabular} }
	\end{center}
\end{table}

Table 3 demonstrates numerical values of the excess Helmholtz free energy obtained by expressions  (\ref{hfe}) and (\ref{hfe_bmcsl}) against the MD values \cite{Heyes18}. In the MS approximation, the ARD\% values for the excess free energy are somewhat larger than those for pressure and excess chemical potential. The BMCSL values are comparable with the MD values, as well.  Note that our all results obtained here are independent on a number of grid points and a length parameter we mentioned above.  
\begin{table}[h]
	\caption{ The same as shown in Table 1, but for the excess Helmholtz free energy.}
	\begin{center}
		{\scriptsize
			\begin{tabular}{c@{\hspace{2mm}}c@{\hspace{2mm}}c@{\hspace{2mm}}c@{\hspace{2mm}}c@{\hspace{2mm}}c@{\hspace{2mm}}c@{\hspace{2mm}}c@{\hspace{2mm}}c@{\hspace{2mm}}c@{\hspace{2mm}}c@{\hspace{2mm}}c@{\hspace{2mm}}c@{\hspace{2mm}}c@{\hspace{2mm}} }
				\hline\hline
				$\eta$ & $\beta a^{e}$ (MS) & $\beta a^{e}$ (BMCSL) & MD~\cite{Heyes18}  \\
				\hline
				0.15 & 0.6 (7.0) & 0.6385 (1.023) & 0.6451 \\
				0.25 & 1.2 (2.9) & 1.2420 (0.4777) & 1.2361 \\
				0.35 & 2.1 (0.3) & 2.0837 (0.487) & 2.0939 \\
				0.45 & 3.6 (8.7) & 3.3298 (0.510) & 3.3129 \\
				\hline\hline
		\end{tabular} }
	\end{center}
\end{table}

After above calculations, we continued our numerical experiments of thermodynamic quantities for a number of binary sets of mixtures, in which one of the three parameters (diameter ratio $\sigma_{2}/\sigma_{1}$, mole fraction of a larger sphere $x_{1}$, and a packing fraction $\eta$) was varied, while the remaining two were kept constant. Numerical values of the MS and BMCSL approaches as a function of $\sigma_{2}/\sigma_{1}$ at $\eta = 0.3$ and $x_{1} = 0.5$ are presented in Figure 1. Numerical values of thermodynamic quantities as functions of $x_{1}$ at $\eta = 0.3$ and $\sigma_{2}/\sigma_{1} = 0.5$ are shown in Figures 2. Figures 3 presents thermodynamic quantities as a function of $\eta$ at $\sigma_{2}/\sigma_{1} = 0.5$ and $x_{1} = 0.5$. In all plots  of Figure 1-3 the cross and solid curve denote the MS and BMCSL results respectively. 
Black point in the plots denotes the MC result \cite{Barosova96}, while blue/red point presents the MD data \cite{Heyes18}.  
\begin{figure}
	\centering
    \mbox{\subfigure{\includegraphics[width=0.4500\textwidth]{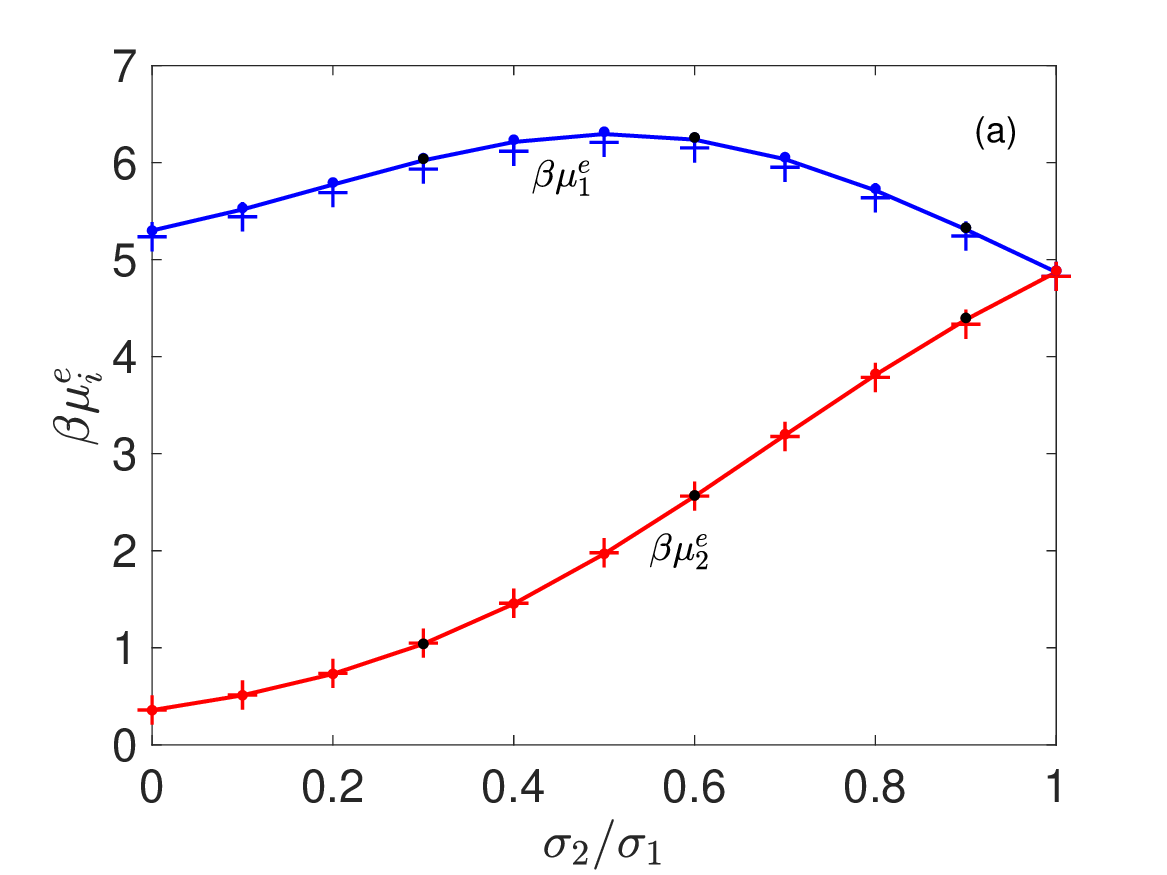}}
               \subfigure{\includegraphics[width=0.4500\textwidth]{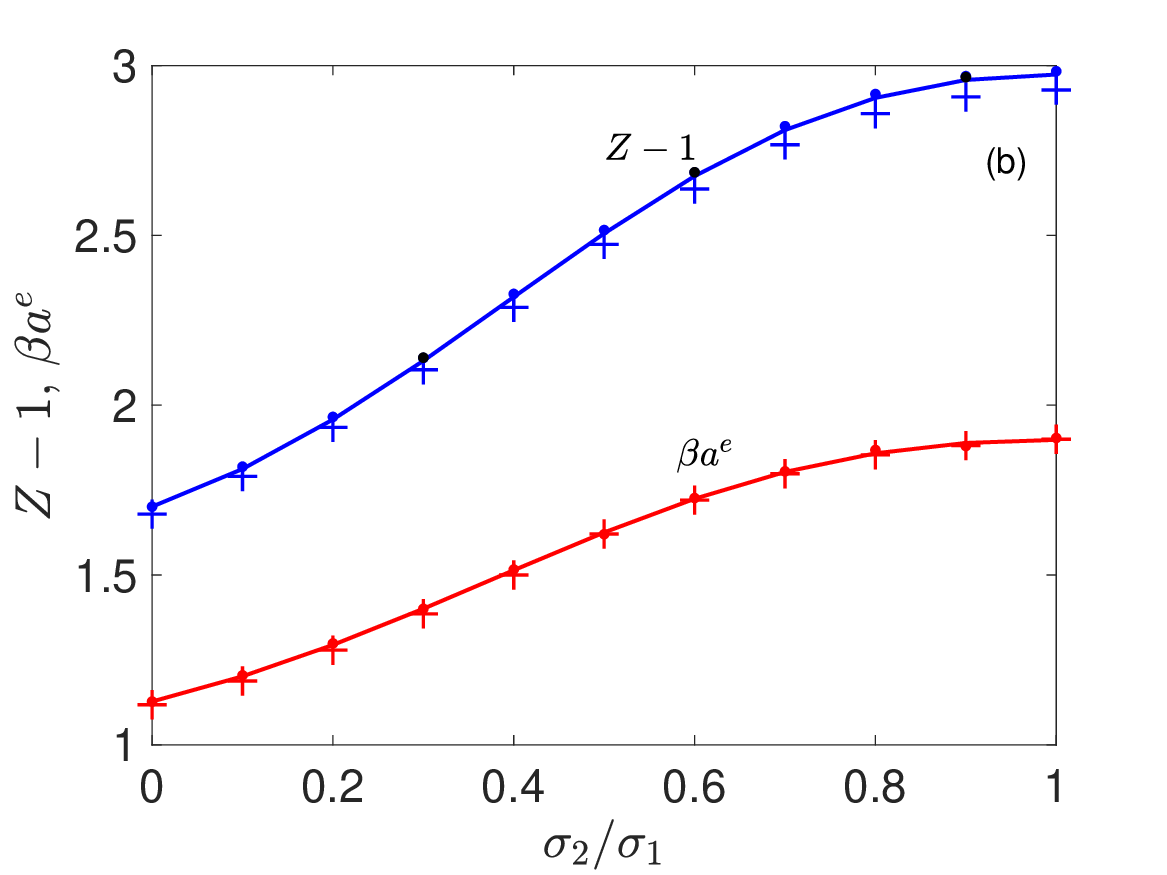}}
      }
		\caption{Results from the MS and BMCSL approaches are shown with crosses and solid curves, respectively. Plots of the excess chemical potential $\beta \mu_{i}$ (a), and the excess Helmholtz free energy (b) and the compressibility factor $Z-1$ (b) as a function of a diameter ratio $\sigma_{2}/\sigma_{1}$ at $\eta  = 0.3$ and $x_{1}=0.5$. Black and blue/red points show the MC~\cite{Barosova96} and MD~\cite{Heyes18} simulations, respectively.} \label{sample-figure}
\end{figure}

From the plots of Figure 1-3, it's been seen that the MD (blue/red points) and BMCSL (solid lines) are almost indistinguishable, and values of excess chemical potential from the MS approach follows more closely the curves than those of the pressure and excess free energy, especially for high values of $\eta$ (Figure 3).    
\begin{figure}
	\centering
	    \mbox{\subfigure{\includegraphics[width=0.4500\textwidth]{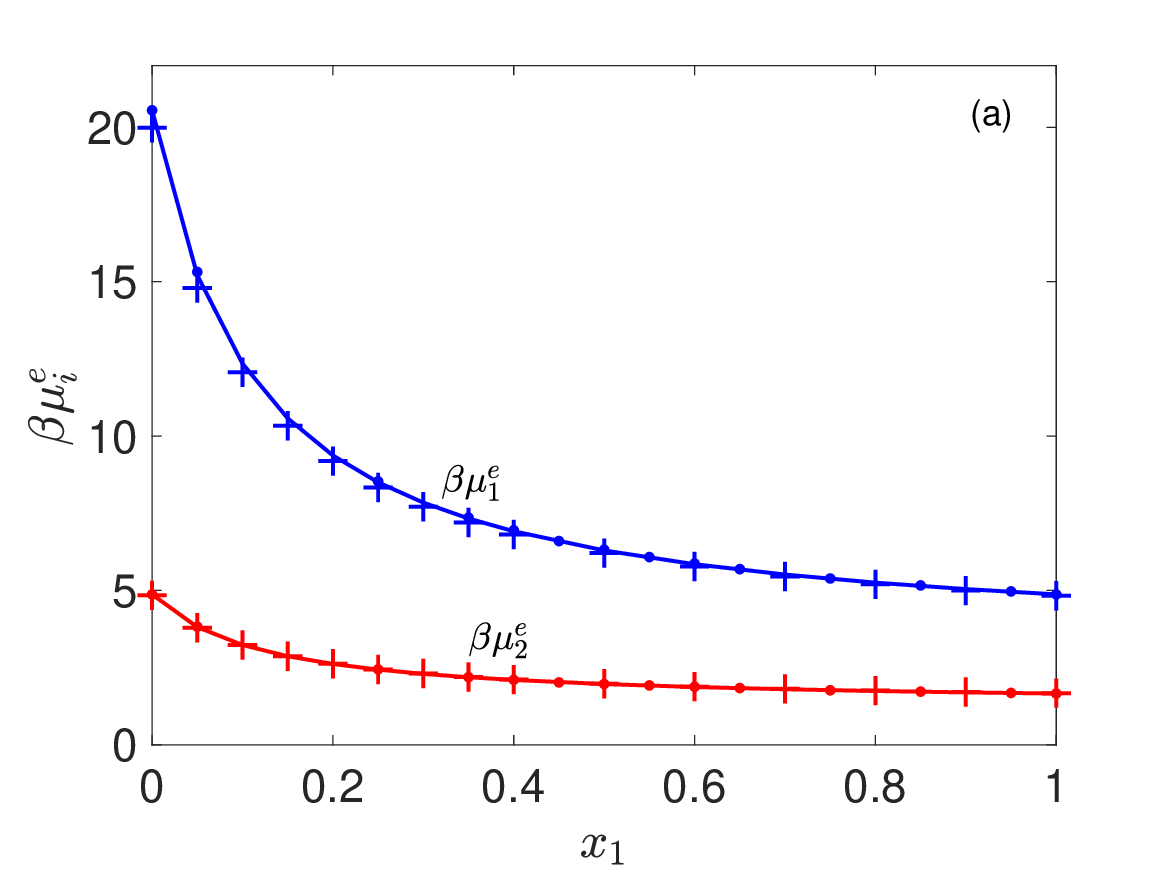}}
                       \subfigure{\includegraphics[width=0.45000\textwidth]{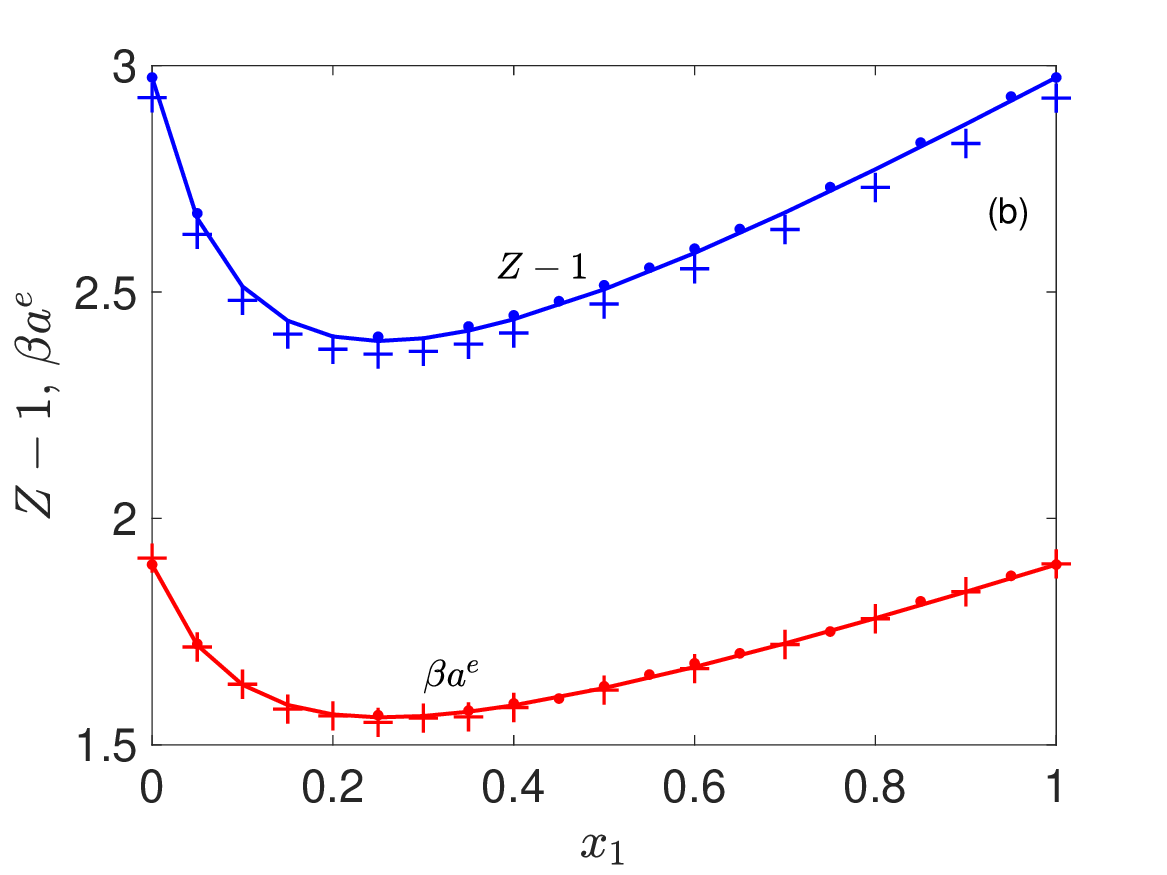}}
			}
	\caption{The same plots as shown in Figure 1, but for as a function of the mole fraction $x_{1}$ at $\eta = 0.30$ and $\sigma_{2}/\sigma_{1} = 0.5$. Points show the MD~\cite{Heyes18} simulations. } \label{sample-figure}
\end{figure}

\begin{figure}
	\centering
    \mbox{\subfigure{\includegraphics[width=0.4500\textwidth]{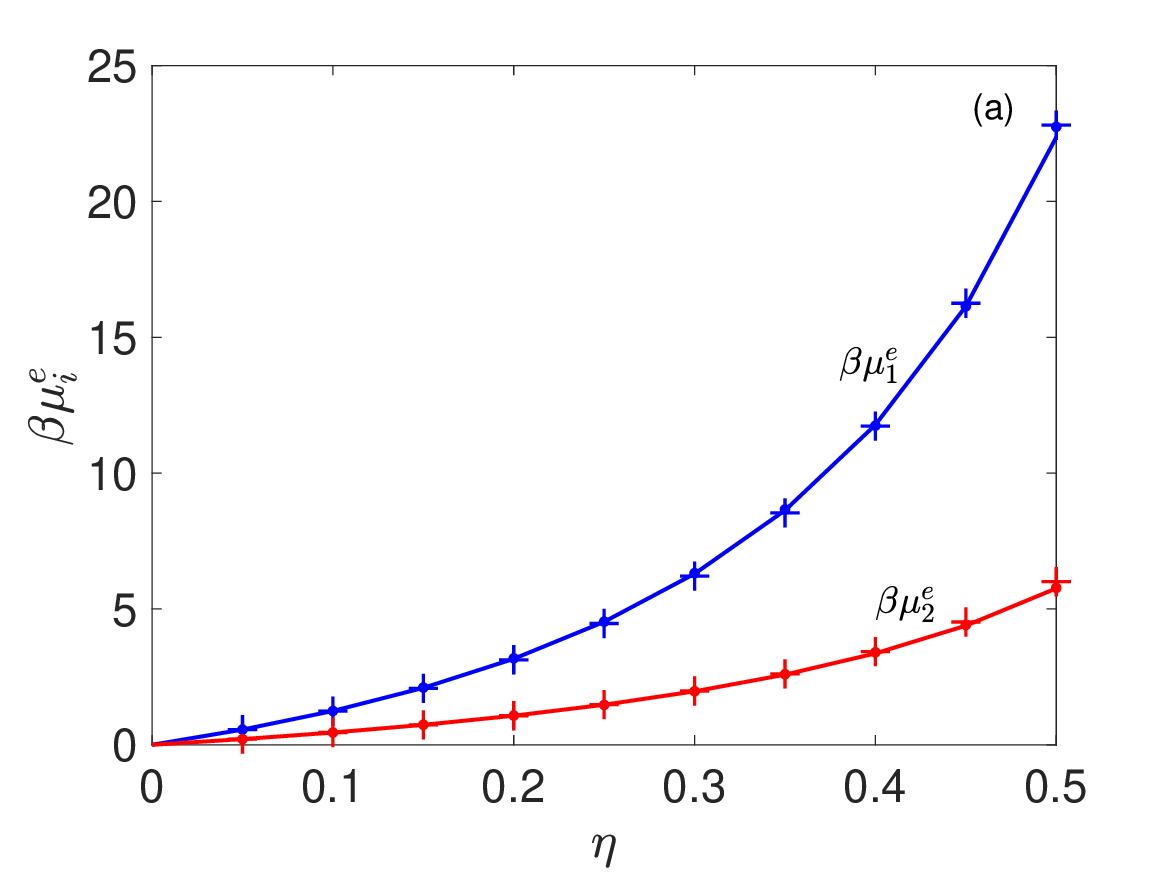}}
               \subfigure{\includegraphics[width=0.4500\textwidth]{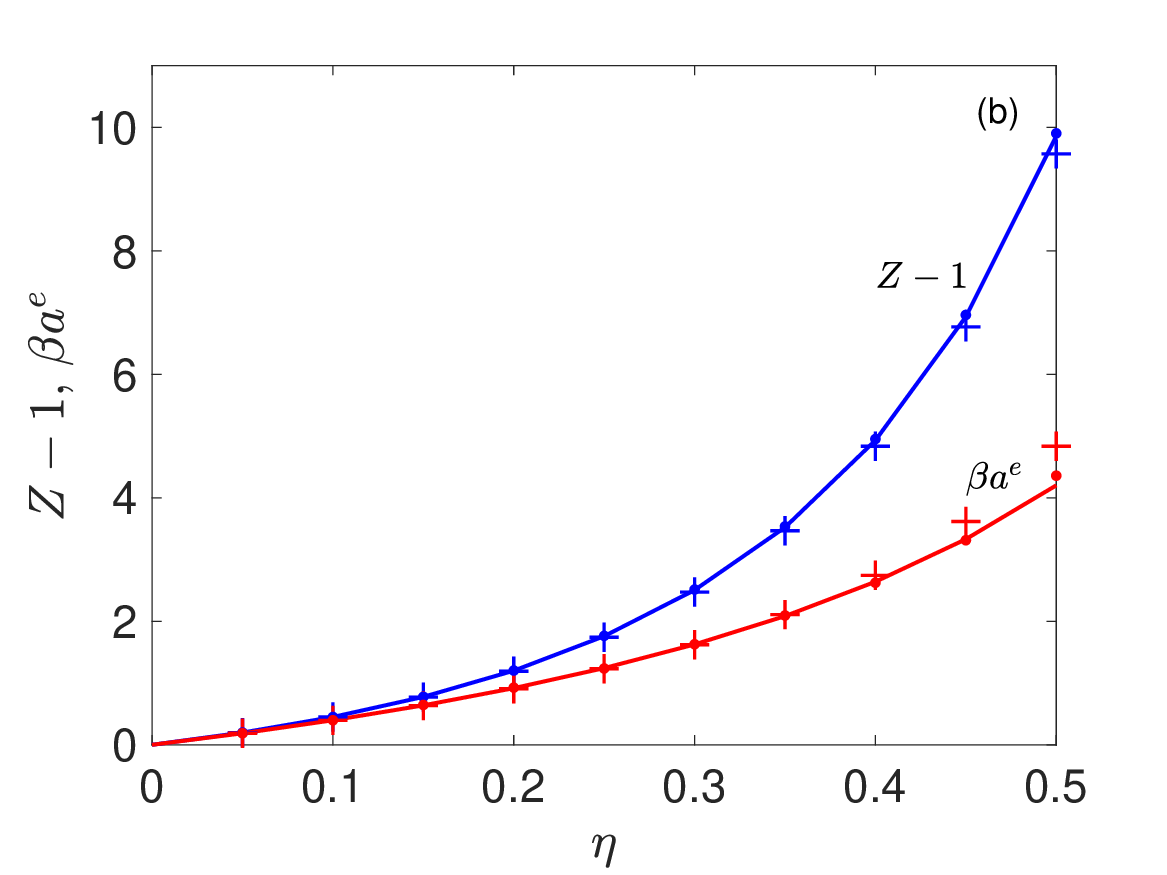}}
		}	
		\caption{The same plots as shown in Figure 1, but for as a function of the packing fraction $\eta$ at $x_{1} = 0.5$ and $\sigma_{2}/\sigma_{1} = 0.5$. Points show the MD~\cite{Heyes18} simulations.} \label{sample-figure}
\end{figure}

In Figure 4 we exhibited similar results for $x_{1} = 0.15$ and $\sigma_{2}/\sigma_{1} = 1.5$ as function of $\eta$, and in this case $\beta \mu_{2}$ takes large values because of large $\sigma_{2}$. 
\begin{figure}
	\centering
    \mbox{\subfigure{\includegraphics[width=0.4500\textwidth]{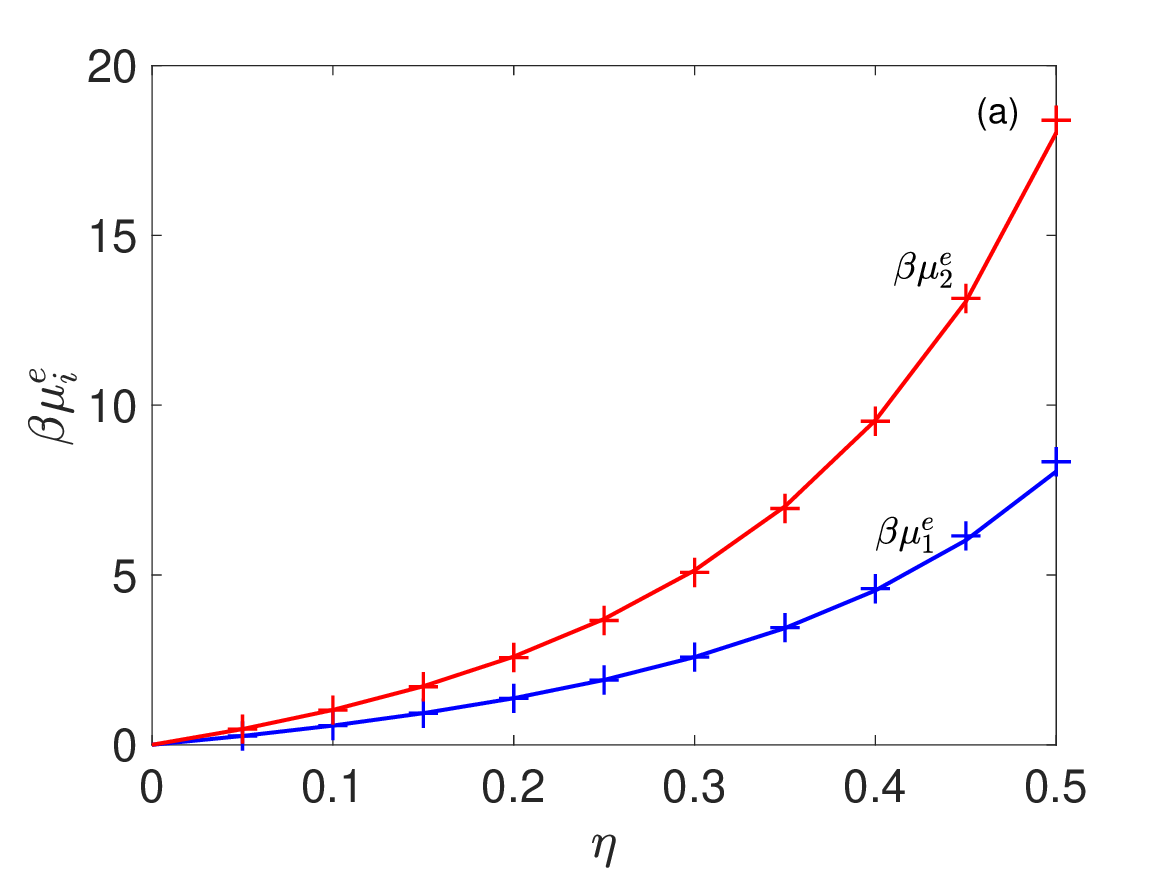}}
               \subfigure{\includegraphics[width=0.4500\textwidth]{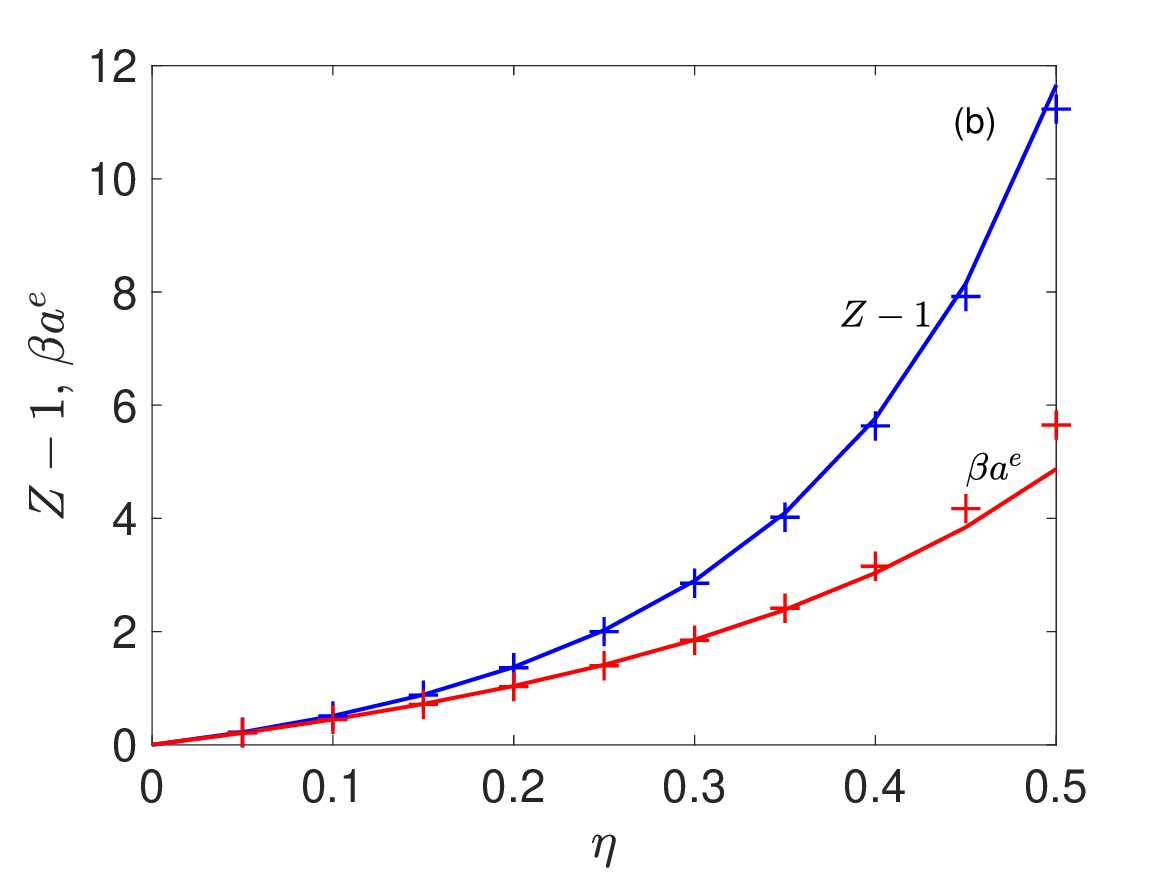}}
		}	
		\caption{The same plots as shown in Figure 3, but for $x_{1} = 0.15$ and $\sigma_{2}/\sigma_{1} = 1.5$. } \label{sample-figure}
\end{figure}

In Figures 5-6, we have shown the ARD\% values for the $\beta \mu^{e}_{1}$ (blue), $\beta \mu^{e}_{2}$ (red), $Z$ (green) and $\beta a^{e}$ (black), which correspond to the plots shown in Figures 1-4 from the MS approximation. The ARD\% values in Figure 5 and Figure 6a are obtained with respect to the MD data \cite{Heyes18}.  From plots in Figure 5, it has been observed that the ARD\% values, which are as a function of $\sigma_{2}/\sigma_{1}$ for $x_{1} = 0.5$ and $\eta = 0.3$, are less than $2\%$, while those, which are as a function of $x_{1}$ for $\sigma_{2}/\sigma_{1} = 0.5$ and $\eta = 0.3$, are less than $3.5\%$.   
\begin{figure}
	\centering
    \mbox{\subfigure{\includegraphics[width=0.4500\textwidth]{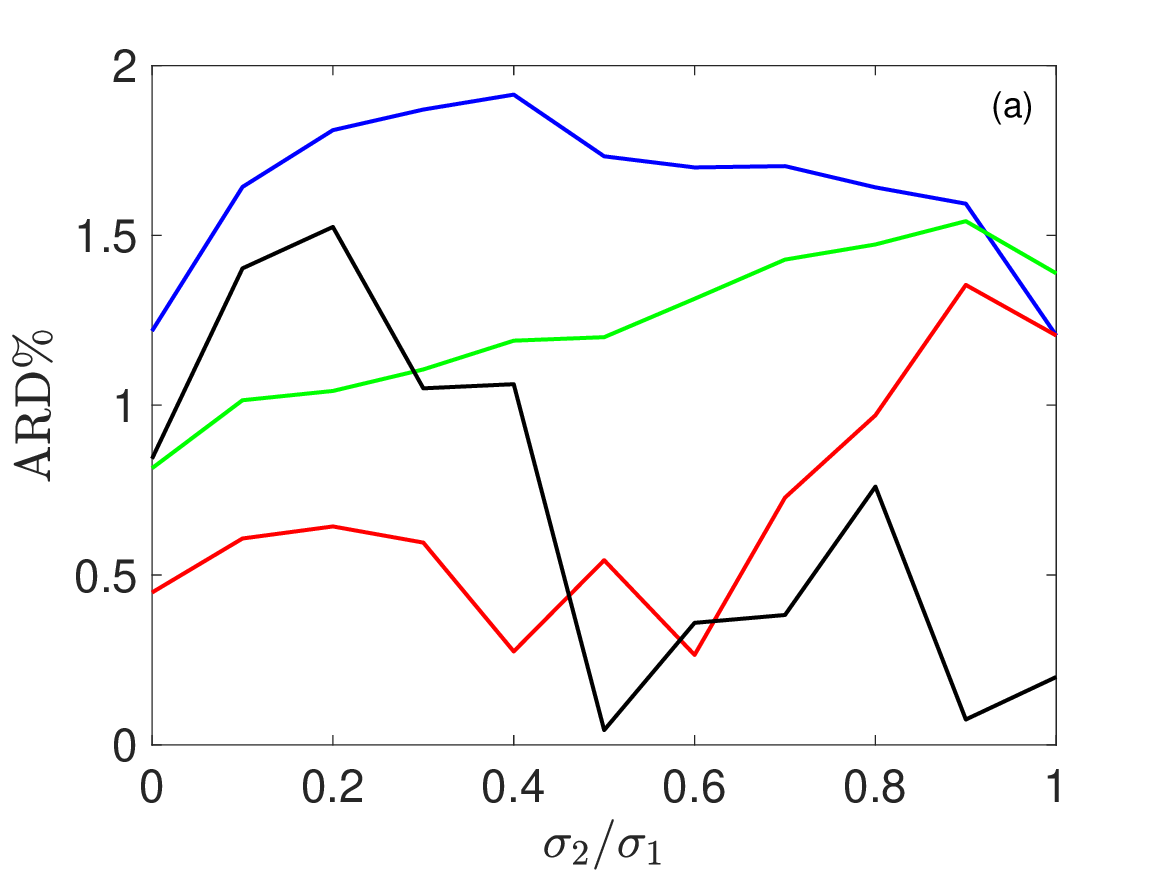}}
               \subfigure{\includegraphics[width=0.4500\textwidth]{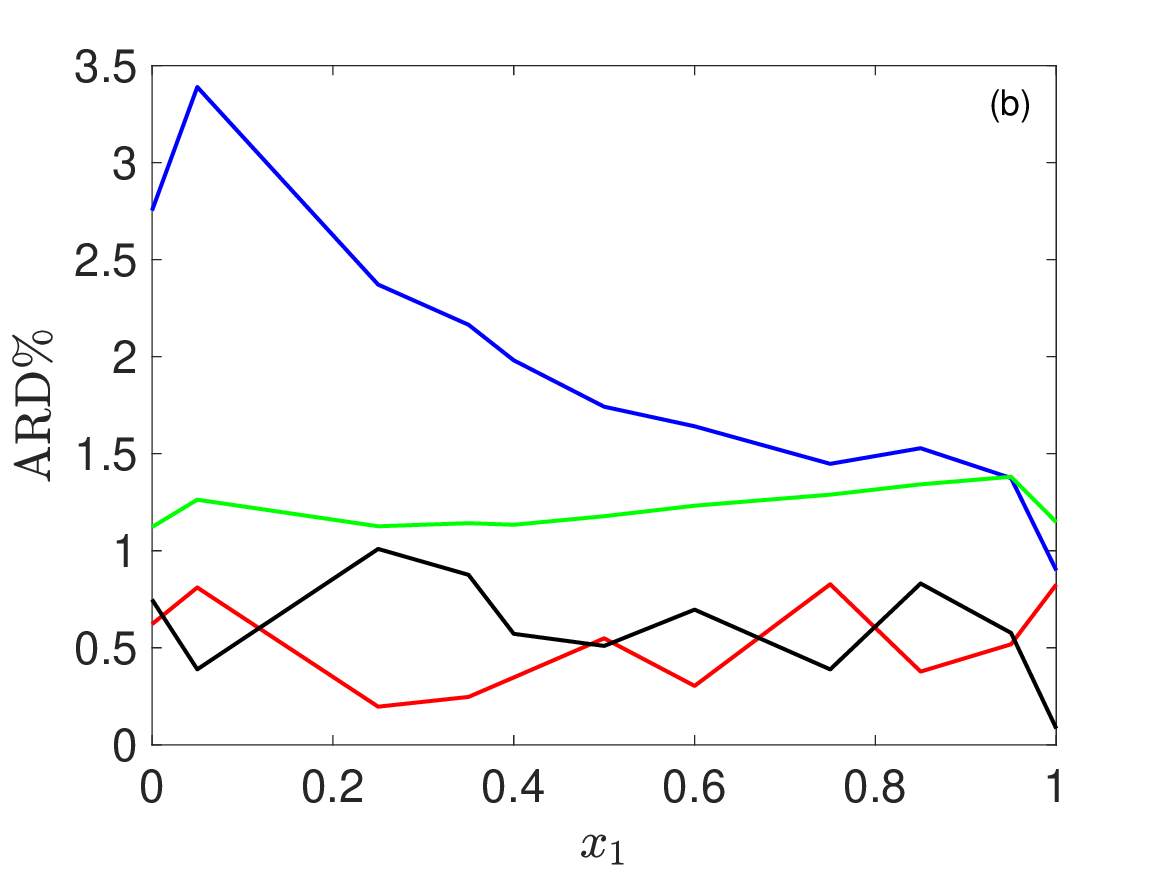}}
		}	
		\caption{(a) Absolute relative deviation percent (ARD\%) of $\beta \mu^{e}_{1}$ (blue),  $\beta \mu^{e}_{2}$ (red),  $Z$ (green), and $\beta a^{e}$ (black) vs $\sigma_{2}/\sigma_{1}$ at $x_{1}=0.5$ and $\eta = 0.3$;  (b) same plots as shown in panel Fig.~5a, but, vs $x_{1}$ at $\sigma_{2}/\sigma_{1}=0.5$ and $\eta = 0.3$. } \label{sample-figure}
\end{figure}

For plots in Figure 6, the ARD\% values of an excess chemical potential and pressure as a function of $\eta$ are less than $5\%$, however, the ARD\% value of an excess free energy becomes large when $\eta > 0.45$, and reaches $\sim16\%$ (black curve in panel Fig.~6b) at $\eta = 0.5$. Note that the ARD\% values for plots in Figure 6b are obtained with respect to the corresponding BMCSL values (curves in Figure 4).  

\begin{figure}
	\centering
    \mbox{\subfigure{\includegraphics[width=0.4500\textwidth]{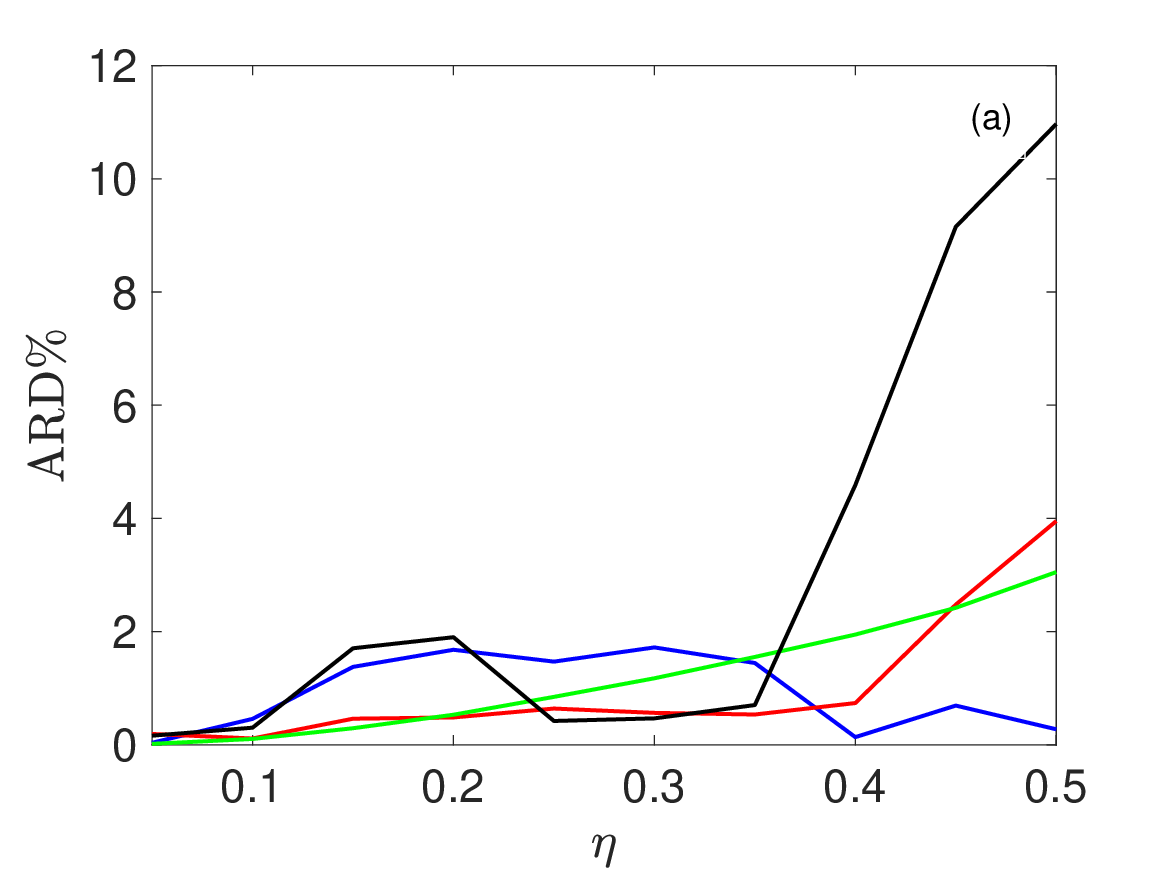}}
               \subfigure{\includegraphics[width=0.4500\textwidth]{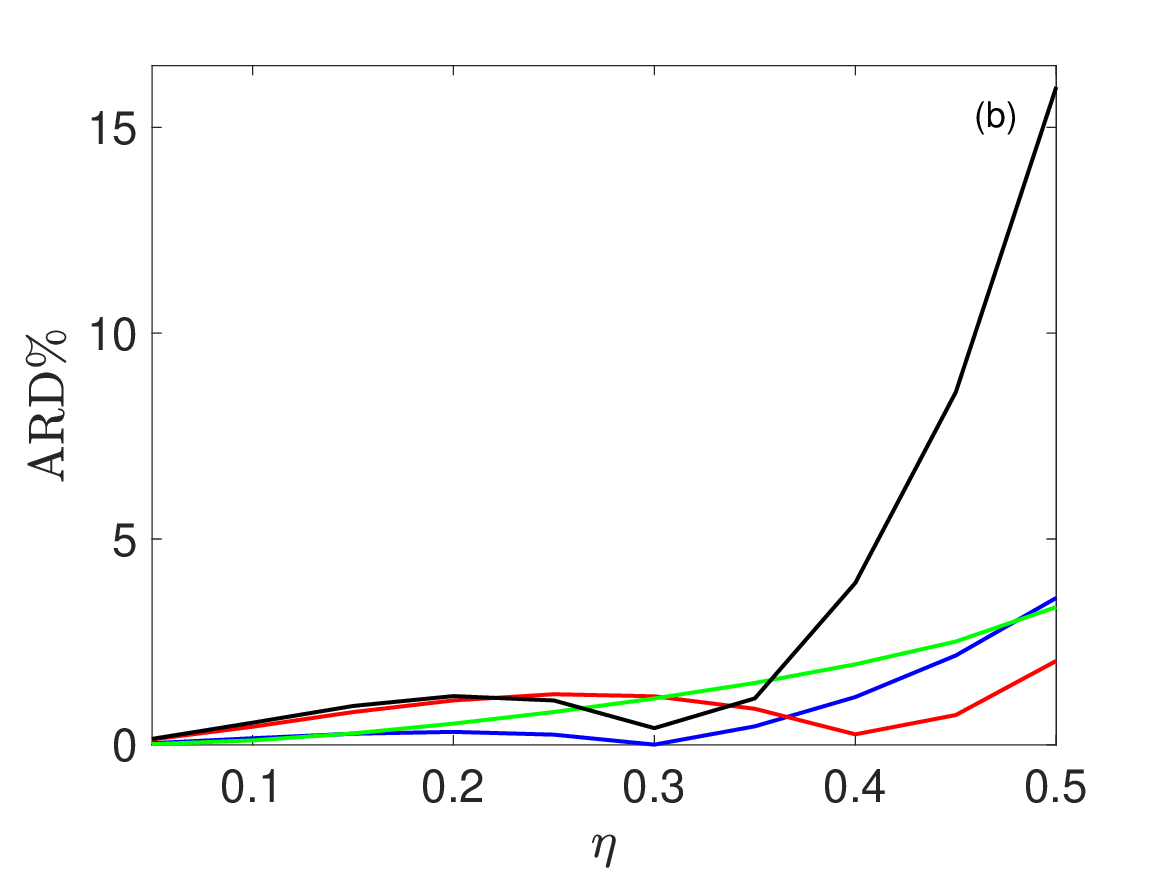}}
		}	
		\caption{(a) Same plots as shown in Fig.5a, but, vs $\eta$ at $x_{1}=\sigma_{2}/\sigma_{1} = 0.5$;  (b) same plots as shown in panel Fig.6a, but, vs $\eta$ at $x_{1} = 0.15$ and $\sigma_{2}/\sigma_{1}=0.5$. } \label{sample-figure}
\end{figure}

Based on plots of the ARD\% values of excess chemical potential in Figures 5 and 6, we note that an approximate expression (\ref{excp}) for evaluating an excess chemical potential can be reliably used for the HS mixture in the MS approximation. Moreover, for the system the virial pressure obtained the MS approximation can be comparable against the accurate BMCSL value or the MD simulation.   

\begin{figure}
	\centering
    \mbox{\subfigure{\includegraphics[width=0.4500\textwidth]{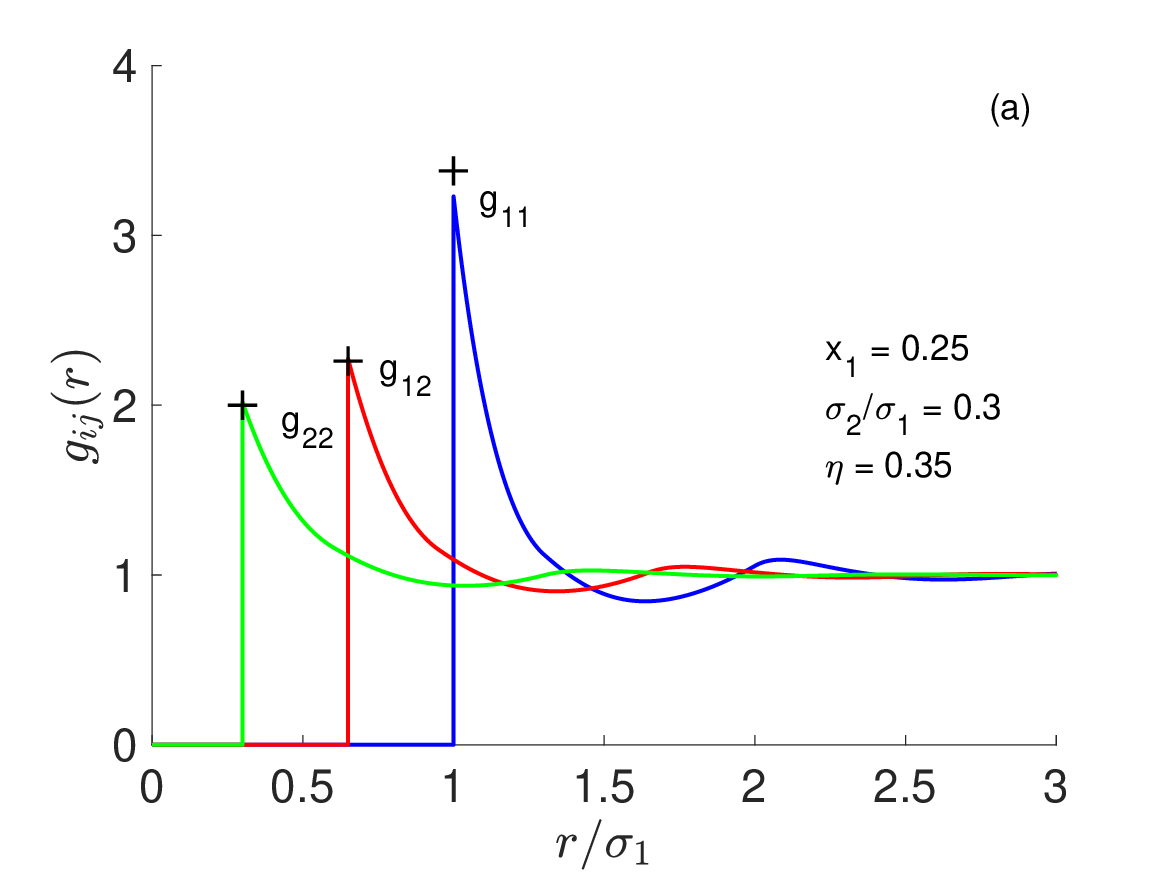}}
               \subfigure{\includegraphics[width=0.4500\textwidth]{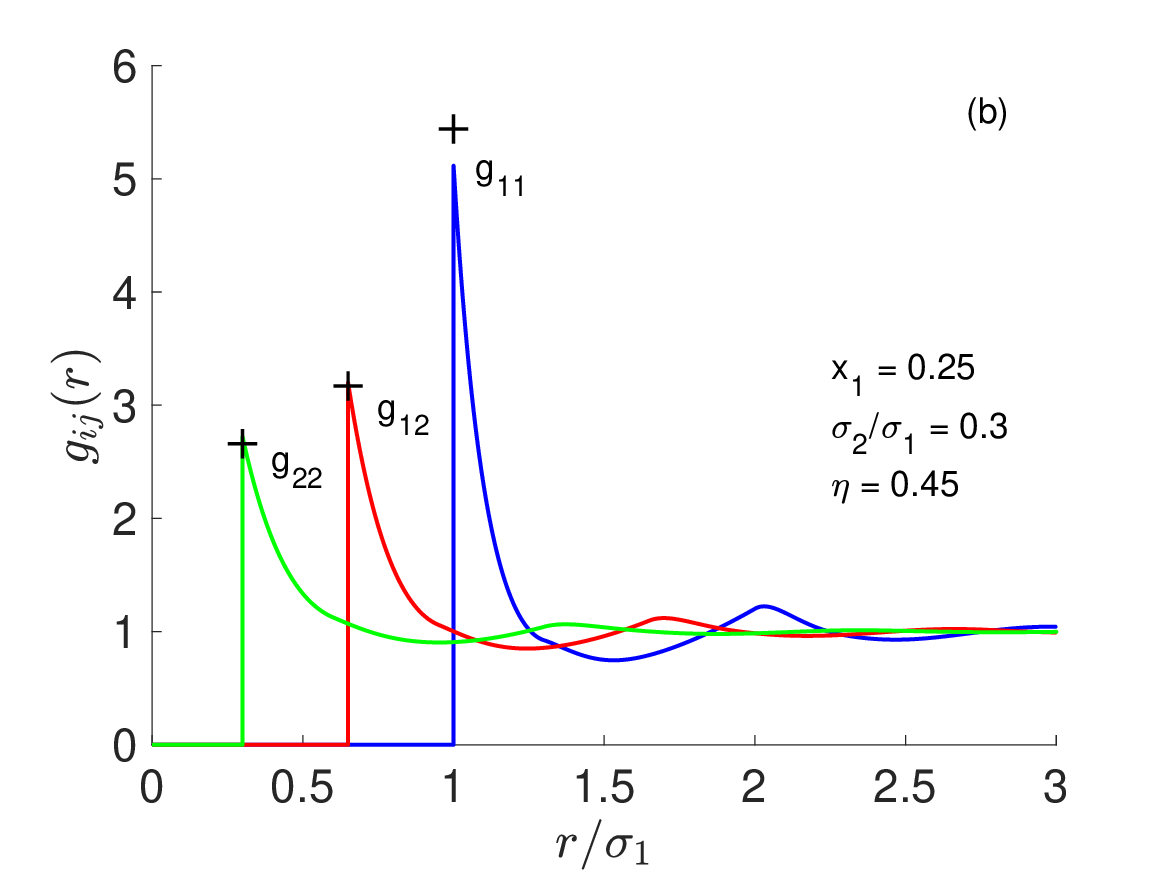}}
		}	
		\caption{Radial distribution functions $g_{ij}$ for the binary HS system for $x_{1} = 0.25$, $\sigma_{2}/\sigma_{1} = 0.3$ and $\eta = 0.35$ (a) and $0.45$ (b).  Cross denoting the contact value is taken from Ref.~\cite{Barosova96}} \label{sample-figure}
\end{figure}

In Figure 7, we have shown the radial distribution function (RDF) for the binary HS mixture at $x_{1} = 0.25, \sigma_{2}/\sigma_{1} = 0.3, \eta = 0.35$ (a) and  $x_{1} = 0.25, \sigma_{2}/\sigma_{1} = 0.3, \eta = 0.45$ (a). The contact values from the MC simulation are taken from Ref.~\cite{Barosova96}, which are very close to the MS data, except for the $g_{11}$ (blue) function.  For plots in panel (a) of Figure 7, the contact values from the MC simulation are $3.38, 2.26$ and $2.00$ for $g_{11}$, $g_{12}$ (red) and $g_{22}$ (green), respectively. However,  their values from the MS approximation are $3.23, 2.27$, and $2.02$. Difference between the contact values for the $g_{11}$ from the MC and MS methods is $0.15$, however, same difference from the MC and PY methods is $0.38$ \cite{Malijevsky97}. For plots in panel (b) of Figure 7, the contact values from the MC simulation are $5.44, 3.17$ and $2.66$ for $g_{11}, g_{12}$ and $g_{22}$, respectively, while they are $5.11, 3.19$ and $2.72$, respectively, from the MS approximation. Difference between the contact values of the $g^{\mathrm{MC}}_{11}$ and $g^{\mathrm{MS}}_{11}$ RDFs is $0.33$, however, similar difference with the $g^{\mathrm{PY}}_{11}$ is $1.01$ \cite{Malijevsky97}. This difference clearly means that the RDF obtained from the MS approximation can be better than that from the PY approximation, which in turn could give better values of thermodynamic quantities, such as, pressure and excess chemical potential. 


\section{Conclusions}	

In this work we have applied the Ornstein-Zernike integral equation for binary additive hard-sphere mixtures at equilibrium using the Martynov-Sarkisov approximation. We computed pressure using the virial route and the BMCSL formula. We obtained an excess chemical potential using an approximate expression based on correlation functions. The excess free energy has been computed using the Euler equation of thermodynamics.  Moreover, excess chemical potential and free energy are evaluated using the BMCSL formulas. Our findings from the MS approximation have been compared with those obtained with the fairly accurate BMCSL formulas and the MD simulation. Furthermore, the absolute relative deviation percent for thermodynamic quantities for the system has been computed as function of diameter ratio of spheres, mole fraction of larger component and packing fraction, in which one of them is varied, while two others are kept constant. Large values of the ARD\% come for the excess free energy for high density, and maximum of it is $\sim 16\%$ at $\eta = 0.5$.    Note that, to our knowledge, this is the first calculation for the thermodynamic properties for the mixture in the MS approximation.


\section*{Acknowledgement(s)}
This research work has been supported by the Mongolian Foundation for Science and Technology (Project No. ShUTBIKhKhZG-2022/167).

\section*{Disclosure statement}
No potential conflict of interest was reported by the authors.

   
\section{Appendix A}
  \renewcommand{\theequation}{A-\arabic{equation}}
  \setcounter{equation}{0}  

The most common technique to obtain the chemical potential $\mu^{e}_{i}$ (i.e., the Gibbs free energy per particle) for the component $i$ in the $N_{m}$ mixtures is through the Kirkwood charging formula~\cite{Kirkwoodjcp35, Kirkwoodjcp36},
\begin{eqnarray}\label{a1}
 \beta \mu_{i} & = & \beta \mu^{id}_{i} + \beta \mu^{e}_{i} \\ \nonumber
  & = & \beta \mu^{id}_{i} + \sum\limits^{N_{m}}_{j=1} \rho_{j} \int^{1}_{0} d\lambda \int d\mathbf{r} \frac{\partial \beta u_{ij}(r, \lambda) }{\partial \lambda} g_{ij}(r,\lambda). 
\end{eqnarray}
where $\beta \mu^{id}_{i}$ and $\beta \mu^{e}_{i}$ are the ideal and excess chemical potentials for the component $i$. In this technique a coupling parameter $\lambda$ scales interactions of one particle added into $N-1$ particle system, that is, when $\lambda = 0$, the particle is removed and when $\lambda = 1$, the particle is fully coupled to the system.  To eliminate a derivative of $\partial u_{i,j}/\partial \lambda$ in the Kirkwood charging formula, we start with an exact expression
\begin{eqnarray}\label{a2}
 [1 + h_{ij}(r, \lambda)] = e^{-\beta u_{ij}(r,\lambda) + h_{ij}(r, \lambda) - c_{ij}(r, \lambda) + B_{ij}(r, \lambda)}.
\end{eqnarray}
Taking the derivatives of both sides, we arrive at
\begin{eqnarray}\label{a3}
  [1 + h_{ij}(r, \lambda)] \frac{\partial \beta u_{ij}(r, \lambda)}{\partial \lambda} & = & \frac{\partial}{\partial \lambda} \Big(\frac{1}{2}h^{2}_{ij}(r) - c_{ij}(r) + B_{ij}(r) \Big) \\ \nonumber
 & & - h_{ij}(r,\lambda)\frac{\partial c_{ij} (r, \lambda)}{\partial \lambda} + h_{ij}(r)\frac{\partial B_{ij} (r,\lambda)}{\partial \lambda}. 
\end{eqnarray}
Inserting an expression (\ref{a3}) into the Kirkwood charging formula for the excess chemical potential (\ref{a1}), we have 
\begin{eqnarray}\label{a4}
 \beta \mu^{e}_{i} & = & \sum\limits^{N_{m}}_{j=1} \rho_{j} \Big[  \int d \mathbf{r} \Big(\frac{1}{2}h^{2}_{ij}(r) - c_{ij}(r) + B_{ij}(r) \Big)  \\ \nonumber
 & & - \int^{1}_{0}d\lambda \int d\mathbf{r} h_{ij}(r,\lambda)\frac{\partial c_{ij} (r,\lambda)}{\partial \lambda} + \int^{1}_{0}d\lambda \int d\mathbf{r} h_{ij}(r,\lambda)\frac{\partial B_{ij} (r,\lambda)}{\partial \lambda}  \Big].
\end{eqnarray}
If we assume that $h_{ij} (r, \lambda) \approx \lambda h_{ij}(r)$, $c_{ij} (r, \lambda) \approx \lambda c_{ij}(r)$, and $B_{ij}(r,\lambda) \approx B_{ij}(\gamma^{2}(r, \lambda))$,  the second and the third integrals of an expression (\ref{a4}) lead to 
\begin{eqnarray}\label{a4a}
\int^{1}_{0}d\lambda \int d\mathbf{r} h_{ij}(r,\lambda)\frac{\partial c_{ij} (r,\lambda)}{\partial \lambda} = \frac{1}{2}\int d\mathbf{r} h_{ij}(r) c_{ij}(r),
\end{eqnarray}
and 
\begin{eqnarray}\label{a4b}
\int^{1}_{0}d\lambda \int d\mathbf{r} h_{ij}(r,\lambda)\frac{\partial B_{ij} (r,\lambda)}{\partial \lambda}   = \frac{2}{3}\int d\mathbf{r} h_{ij}(r) B_{ij}(r),
\end{eqnarray}
respectively. 

Combining the expressions (\ref{a4}), (\ref{a4a}) and (\ref{a4b}), we have
\begin{eqnarray}\label{a5}
 \beta \mu^{e}_{i} & = & \sum\limits^{N_{m}}_{j=1} \rho_{j}\int d \mathbf{r}  \Big[  \Big(\frac{1}{2}h^{2}_{ij} - c_{ij}- \frac{1}{2} h_{ij} c_{ij}\Big) + \Big(1+ \frac{2}{3} h_{ij} \Big) B_{ij}\Big].
\end{eqnarray}


\end{document}